\documentclass[sigconf, nonacm]{acmart}

\newcommand\vldbdoi{XX.XX/XXX.XX}
\newcommand\vldbpages{XXX-XXX}
\newcommand\vldbvolume{15}
\newcommand\vldbissue{1}
\newcommand\vldbyear{2022}
\newcommand\vldbauthors{\authors}
\newcommand\vldbtitle{\shorttitle} 

\newcommand\vldbpagestyle{plain} 

\usepackage{balance} 
\usepackage{amsmath,epsfig}
\usepackage{graphicx}
\usepackage{booktabs}
\usepackage{mathtools,mathrsfs}
\usepackage{paralist}
\usepackage{pifont}
\usepackage{listings}
\usepackage{epstopdf}
\usepackage{xspace,colortbl,subfigure,xcolor}
\usepackage{bm,bbold,multirow}

\newcommand{\llib}{\textsf{LLib}\xspace}

\newcommand{\qr}[1]{\textit{Query #1}}
\newcommand{\rasql}{{\sc RaSQL}\xspace}

\newcommand{\datalog}{{\sc Datalog}\xspace}

\newcommand{\mlib}{\textsf{MLlib}\xspace}

\newcommand{\bldl}{\smallskip\[\begin{array}{ll}}
\newcommand{\cldl}{\[\begin{array}{lrcl}}
\newcommand{\eldl}{\end{array}\]\rm}
\newcommand{\prule}[3]{ #1 & \mt #2 & \leftarrow & \mt #3 \\}

\def\pbody#1#2{ & \mt #1 & & \mt #2 \\}
\newcommand{\rij}[2]{\ensuremath{\textsf{r}_{#1,#2}}\xspace}

\let\STATEXtilde=\~
\renewcommand*{\~}{\relax\ifmmode\expandafter\widetilde\else\expandafter\STATEXtilde\fi}
\newcolumntype{B}{>{\centering\arraybackslash}m{1.5cm}}
\newcolumntype{D}{>{\centering\arraybackslash}m{2.5cm}}

\def\mt{\tt}

\AtBeginDocument{%
	\providecommand\BibTeX{{%
			\normalfont B\kern-0.5em{\scshape i\kern-0.25em b}\kern-0.8em\TeX}}}
\begin{document}
	\title{Demonstration of LogicLib: An Expressive Multi-Language Interface over Scalable Datalog System}
	\author{Mingda Li}
	\email{limingda@cs.ucla.edu}
	\affiliation{%
		\institution{UCLA}
		\country{United States}
	}
	\author{Jin Wang}
	\email{jinwang@cs.ucla.edu}
	\affiliation{%
		\institution{UCLA}
		\country{United States}
	}
	\author{Guorui Xiao}
	\email{grxiao@ucla.edu}
	\affiliation{%
		\institution{UCLA}
		\country{United States}
	}
	\author{Youfu Li}
	\email{youfuli@cs.ucla.edu}
	\affiliation{%
		\institution{UCLA}
		\country{United States}
	}
	\author{Carlo Zaniolo}
	\email{zaniolo@cs.ucla.edu}
	\affiliation{%
		\institution{UCLA}
		\country{United States}
	}
	
	\renewcommand{\shortauthors}{Li et al.}

\begin{abstract}
	With the ever-increasing volume of data, there is an urgent need to provide expressive and efficient tools to support Big Data analytics.
	The declarative logical language Datalog has proven very effective at expressing concisely graph, machine learning, and knowledge discovery applications via recursive queries. 
	In this demonstration, we develop Logic Library (\llib), a library of recursive algorithms written in Datalog that can be executed in BigDatalog, a Datalog engine on top of Apache Spark developed by us.
	\llib encapsulates complex logic-based algorithms into high-level APIs, which simplify the development and  provide a unified interface akin to the one of Spark MLlib. 
	As \llib is fully compatible with DataFrame, it enables the integrated utilization of its built-in applications and new Datalog queries with existing Spark functions, such as those provided by \mlib and Spark SQL. 
	With a variety of examples, we will (i) show how to write programs with \llib to express a variety of applications; (ii) illustrate its user experience in Apache Spark ecosystem; and (iii) present a user-friendly interface to interact with the \llib framework and monitor the query results.
\end{abstract}

	\maketitle
	\pagestyle{\vldbpagestyle}
	\begingroup\small\noindent\raggedright\textbf{PVLDB Reference Format:}\\
	\vldbauthors. \vldbtitle. PVLDB, \vldbvolume(\vldbissue): \vldbpages, \vldbyear.\\
	\href{https://doi.org/\vldbdoi}{doi:\vldbdoi}
	\endgroup
	\begingroup
	\renewcommand\thefootnote{}\footnote{\noindent
		This work is licensed under the Creative Commons BY-NC-ND 4.0 International License. Visit \url{https://creativecommons.org/licenses/by-nc-nd/4.0/} to view a copy of this license. For any use beyond those covered by this license, obtain permission by emailing \href{mailto:info@vldb.org}{info@vldb.org}. Copyright is held by the owner/author(s). Publication rights licensed to the VLDB Endowment. \\
		\raggedright Proceedings of the VLDB Endowment, Vol. \vldbvolume, No. \vldbissue\ %
		ISSN 2150-8097. \\
		\href{https://doi.org/\vldbdoi}{doi:\vldbdoi} \\
	}\addtocounter{footnote}{-1}\endgroup

\section{Introduction}\label{sec-intro}

In the past years, there is a resurgence of Datalog due to its ability to specify declarative data-intensive applications that execute efficiently over different systems and architectures~\cite{DBLP:journals/debu/ZanioloD0LL021}.
The recent theoretical advances~\cite{CompletedAggregates,DBLP:journals/tplp/ZanioloYDSCI17,DBLP:journals/vldb/MazuranSZ13} enable the usage of aggregates in recursions, and this leads to considerable improvements in the expressive power of Datalog.
In response to supporting Datalog queries over large-scale of datasets, many parallel and distributed Datalog engines have been developed by researchers from both academia and industry.
Following this line of efforts, we have proposed \textsf{BigDatalog}~\cite{DBLP:conf/sigmod/ShkapskyYICCZ16}, a Datalog engine on top of Apache Spark~\cite{DBLP:conf/nsdi/ZahariaCDDMMFSS12}.
Similar with the rich libraries provide by Spark such as Spark SQL, GraphX, MLlib and SparkR, \textsf{BigDatalog} provides the query interface of Datalog which can benefit from the efficient in-memory computation for analytical workloads with complicated recursions. 

On the basis of above efforts for Datalog system, we further develop several advanced applications upon it. 
\textsf{DatalogML}~\cite{DBLP:journals/vldb/WangWLGDZ21} supports training different machine learning models with gradient descent written in Datalog queries by enabling mutual and non-linear recursion in \textsf{BigDatalog}.
\textsf{KDDLog}~\cite{DBLP:conf/icde/LiWLD0Z21} aims at developing data mining algorithms such as frequent item set mining, clustering and decision tree in Datalog.
Both of them have shown superior performance than the original machine learning library \mlib in Apache Spark.
Except for Datalog, we also build the \rasql~\cite{DBLP:conf/sigmod/0001WMSYDZ19,DBLP:conf/sigmod/WangX0WZ20} language and system to enable expressing aggregating in recursive SQL queries.
However, it requires to develop more easy-to-use APIs as that of \mlib so as to make them compatible to the Apache Spark ecosystem and used by broader audience.

In this demonstration, we propose \textsf{Logic Library} (\llib), a library of applications written in Datalog programs built on top of the \textsf{BigDatalog} system.
To be more specific, \llib not only includes popular recursive algorithms like the graph search ones~\cite{iclp19das}, but also machine learning and data mining algorithms introduced in~\cite{DBLP:journals/vldb/WangWLGDZ21} and~\cite{DBLP:conf/icde/LiWLD0Z21}  that is supported by \mlib now.
It provides a unified interface that is similar with \mlib so that it is rather easy for the audience of Spark community to use it.
Users can utilize the built-in functions that encapsulate Datalog programs to express a variety of applications by using the DataFrame APIs.
\llib also supports APIs in multiple programming languages such as Scala, Java and Python.
To sum up, the \llib framework enables users to use the highly efficient Datalog applications in the similar way of those in Apache Spark \mlib thus significantly improve the usability of Datalog based systems.

\section{The Logic Lib Framework}\label{sec-llib}
\begin{figure*}[h]
	\centering
	\vspace{-1em}
	\includegraphics[width=0.86\textwidth]{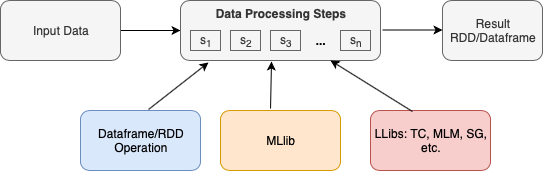}
	\vspace{-.5em}
	\caption{The Overall Architecture}\label{fig:framework}
\end{figure*}
\subsection{The BigDatalog System}\label{subsec-bg}

As introduced above, the \llib framework is built based on \textsf{BigDatalog}~\cite{DBLP:conf/sigmod/ShkapskyYICCZ16}, which is a distributed Datalog engine on top of Apache Spark.
In this section, we first give a brief introduction of the system.

\textsf{BigDatalog} supports relational algebra, aggregation and recursion, as well as a host of declarative optimizations.
It uses and extends Spark SQL~\cite{DBLP:conf/sigmod/ArmbrustXLHLBMK15} operators, and also introduces a novel recursive operator implemented in the Catalyst framework so that its planning features can be used on the recursive plans of Datalog programs.
\textsf{BigDatalog} can resolve recursion in the compilation step by recognizing recursive tables when building the operator tree and evaluate the recursive program using the semi-naive method~\cite{DBLP:conf/datalog/ShawKHS12}.

By enabling non-linear and mutual recursion, we can support expressing more complicated applications like machine learning~\cite{DBLP:journals/vldb/WangWLGDZ21} and data mining~\cite{DBLP:conf/icde/LiWLD0Z21} using \datalog on the above systems.

\subsection{Overview of \llib}\label{subsec-overall}

The overall architecture of \llib is shown in Figure~\ref{fig:framework}.
Similar with that of \mlib, we encapsulate applications written in Datalog programs into functions which can be imported and called in programs written in high-level programming languages.
As a result, these can be seamlessly integrated with other features of the Apache Spark ecosystem.
For example, in an \llib application, there can be steps consisting of regular DataFrame operations, the data transformations from \mlib and functions written in the Datalog provided by \llib.
Within a pipeline illustrated in Figure~\ref{fig:framework}, all steps from preparing Datalog input data to persisting and loading Datalog execution results for subsequent processing can be sequentially executed without extra programming efforts.  

Similar with that of \mlib, \llib also supports the interfaces for multiple programming languages, i.e. Java, Scala and Python.
Since Java is interpretable with Scala, it is relatively simple to support Java interface. 
The remaining gaps between Scala and Java version are mainly the conversions of data collections. 
We bridge these gaps using collections known by both languages in Scala implementation and using the implicit converting mechanism in Scala.
To support Python interface, we utilize Py4J~\footnote{https://www.py4j.org/index.html}, a bridge between Java and Python language, and design a Removal and Recovery mechanism where the transferred DataFrame in PySpark will be converted to a common data collection type acceptable by both Java and Python languages. 
Then during execution in Java, we could infer the schema or directly get the schema from users.

\subsection{Programing with \llib}\label{subsec-steps}

Next we will describe how to write a program with built-in functions in \llib through the example of Transitive Closure. 

\subsubsection{Working Session and Acquiring Data}

In \llib, the first step is to construct a working environment for the Datalog queries and libraries. 
We follow the paradigm of building Spark Session in a similar way as that shown Figure~\ref{fig:worksession}.

\begin{figure}[h!t]\vspace{-1em}
	
	\lstset
	{ 
		basicstyle=\footnotesize,
		numbers=left,
		stepnumber=1,
		keywordstyle=\color{brown},
		xleftmargin=0.5cm,
		showstringspaces=false,
		tabsize=1,
		breaklines=true,
		breakatwhitespace=false,
	}
	
	\begin{lstlisting} 
	session = LogiclibSession.builder().appName("TC")\\
	.master("local[*]").getOrCreate()
	schema = StructType(List(StructField("Point1", IntegerType, true),
	StructField("Point2", IntegerType, true)))
	df = session.read.format("csv").option("header","false").schema(schema).load("arc.csv")
	
	\end{lstlisting}
	\vspace{-.5em}
	\caption{Snippet Code for Working Session and Acquiring Data}
	\label{fig:worksession}
\end{figure}

Here \emph{TC} is a built-in function of \llib for Transitive Closure written in Datalog shown in Figure~\ref{fig:tc}.
\emph{LogiclibSession} synthesizes the Spark environment with required designs for logical programs. 
Within the same session, users can use both Spark libraries like data loading functions to create a DataFrame \textit{df} (line 4) and the functions in our \llib framework. 

\begin{figure}[h]\vspace{-1em}
	
	\lstset
	{ 
		basicstyle=\footnotesize,
		numbers=left,
		stepnumber=1,
		keywordstyle=\color{brown},
		xleftmargin=0.5cm,
		showstringspaces=false,
		tabsize=1,
		breaklines=true,
		breakatwhitespace=false,
	}
	
	\begin{lstlisting} 
	database({
	arc(From: integer, To: integer)
	}).
	tc(From,To)<- arc(From,To).
	tc(From,To) <- tc(From,Tmp), arc(Tmp,To). 
	query tc(From, To).
	\end{lstlisting}
	\vspace{-.5em}
	\caption{Query 1, Rules for Transitive Closure}
	\label{fig:tc}
\end{figure}

\subsubsection{Initializing an Executable Object and Schema Mapping}

In \llib, the data processing pipeline for a typical application is wrapped into an executable object, for which users control initialization and  parameter setting.
We initialize a transitive closure object \emph{tc} with the TC function. 
Then, we can set the property in the way shown in Figure~\ref{fig:tcinitial}.

\begin{figure}[h!t]\vspace{-1em}
	
	\lstset
	{ 
		basicstyle=\footnotesize,
		numbers=left,
		stepnumber=1,
		keywordstyle=\color{brown},
		xleftmargin=0.5cm,
		showstringspaces=false,
		tabsize=1,
		breaklines=true,
		breakatwhitespace=false,
	}
	
	\begin{lstlisting} 
	import edu.ucla.cs.wis.bigdatalog.spark.Logiclib.TC 
	val tc = new TC() 
	tc.setDirection(FromCol = "Node1", ToCol = "Node2")
	\end{lstlisting}
	\vspace{-.5em}
	\caption{Transitive Closure Initialization}
	\label{fig:tcinitial}
\end{figure}

Among these built-in functions, all the libraries need the function \textit{setDirection}, an extension to existing DataFrame made by us, for schema mapping. 
One attribute may have different names in different DataFrames. 
With schema mapping, we could know the corresponding relationship between input DataFrames and the attributes in \llib.
For example, in Query 1 shown in Figure~\ref{fig:tc} there are two attributes \emph{From} and \emph{To} in the \emph{arc} table. 
They can be called in different ways such as (Node1, Node2) as shown in Figure~\ref{fig:tcinitial}. 
Then the mapping between (From, To) and (Node1, Node2) is provided by the \emph{setDirection} function. 


\subsubsection{Execution and Persistence}

With the executable object and imported data, the execution and persistence can be merely a one-line execution with a pre-defined function (\emph{run, genDF} or \emph{genRDD}) defined by us. 
These three functions are implemented for each library in \llib. 
They expect the input data (Dataframe) and the environment (Session) as inputs. 
Their functionality satisfies all basic requirements needed to operate on data and preserve the result as a variable or a file. 
As shown in Figure~\ref{fig:tcllib}, the function \emph{run} executes the logical programs and persist the result directly to the target address. 
It can also output the data into a DataFrame or RDD as shown in second and third line of Figure~\ref{fig:tcllib}. 
In this way, users can also apply other APIs provided by Apache Spark for further post-processing or execution.

\begin{figure}[h]\vspace{-1em}
	
	\lstset
	{ 
		basicstyle=\footnotesize,
		numbers=left,
		stepnumber=1,
		keywordstyle=\color{brown},
		xleftmargin=0.5cm,
		showstringspaces=false,
		tabsize=1,
		breaklines=true,
		breakatwhitespace=false,
	}
	
	\begin{lstlisting} 
	tc.run(df, output = "File", session) 
	val dfNew = tc.genDF(df, session) 
	val rddNew = tc.genRDD(df, session)
	\end{lstlisting}
	\vspace{-.5em}
	\caption{Execution of Transitive Closure with \llib}
	\label{fig:tcllib}
\end{figure}\vspace{-1em}

\section{Applications}\label{sec-app}

In this section, we introduce two typical applications of \llib.
Due to the space limitation, we omit the full Datalog programs here and just provide the references for them, respectively.
\subsection{Recursive Algorithm}\label{subsec-rec}

We first show how to support a typical recursive algorithm in \llib.
Here we use the Datalog program for Multi-level Market (MLM) Bonus~\footnote{https://en.wikipedia.org/wiki/Multi-level\_marketing.} as an example.
In the organization, new members are recruited by and get products from old members (sponsors). One member's bonus is based on his own sales and a proportion of the sales from the people directly or indirectly recruited by him. The  scale of the proportion is user-defined. 
There are two relations in MLM Bonus, including the \emph{sponsor} and \emph{sales}. 
The sponsor relation describes the recruiting relationship among members, while the sales relation records the profits for each member.
In Datalog syntax, the base case should be calculating the member's bonus by the sales table. 
And the recursive rule is to calculate the bonus based on the basic profits and the profits derived from the downstream members.

Under the \llib framework, users could implement this program by a built-in function \emph{MLM}.
The full program of \emph{MLM} can be found in Program 10 of~\cite{DBLP:conf/sigmod/ShkapskyYICCZ16}.
The program using \llib is detailed in Figure~\ref{fig:mlmllib}.

\begin{figure}[ht]\vspace{-1em}
	
	\lstset
	{ 
		basicstyle=\footnotesize,
		numbers=left,
		stepnumber=1,
		keywordstyle=\color{brown},
		xleftmargin=0.5cm,
		showstringspaces=false,
		tabsize=1,
		breaklines=true,
		breakatwhitespace=false,
	}
	
	\begin{lstlisting} 
	val appMLM = new MLM()
	appMLM.setDirection(MCol = "Member", ProfitCol = "Bonus")
	appMLM.setSecDirection(MCol = "Mem1", M2Col = "Mem2")
	appMLM.run(Array(Sales, Sponsor), output = "result", session) 
	\end{lstlisting}\vspace{-.5em}
	\caption{MLM in \llib}\vspace{-1em}
	\label{fig:mlmllib}
\end{figure}

Here we start with two relations \emph{Sales} and \emph{Sponsor} stored in DataFrame. 
We first build an executable object of MLM and then set the schema mappings for two relations with functions \emph{setDirection} and \emph{setSecDirection}, which is a DataFrame function defined by us that is similar with \emph{setDirection}.
To operate the data and persist to \emph{resMLM} file, we use the \emph{run} function in the 4-th line.  

Since most graph algorithms can be also easily expressed with Datalog in a similar way~\cite{DBLP:journals/tplp/CondieDISYZ18}, the \llib framework can support them in a similar way.

\subsection{Machine Learning and Data Mining}\label{subsec-ml}
\begin{figure}[ht]
	\centering
	\vspace{-1em}
	\includegraphics[width=0.4\textwidth]{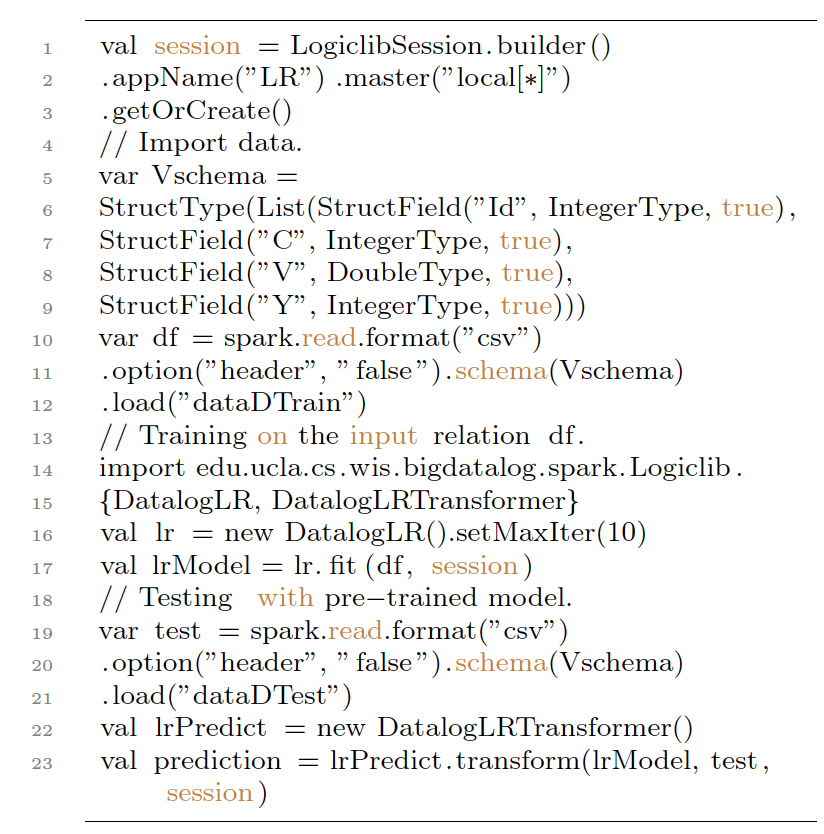}
	\vspace{-.5em}
	\caption{Snippet Code for \llib: Logistic Regression}\label{fig:expLR}\vspace{-1em}
\end{figure}

Then we show how to express machine learning applications with \llib.
The example in Figure~\ref{fig:expLR} expresses the process of training a Logistic Regression classifier on the training data \textsf{dataDTrain}, and making prediction on the test data, \textsf{dataDTest}.  
Here \textsf{DatalogLR} is the Datalog program for logistic regression which can be found in Query~3 of~\cite{DBLP:journals/vldb/WangWLGDZ21}.
To make use of the \datalog programs for machine learning, we first construct a working environment, i.e. \textsf{LogiclibSession}, for our library of machine learning algorithm (line:~1 to~3). 
After importing the required training and predicting functions for Logistic Regression (line:~14 to~15), we can build executable objects for training \textsf{lr} (line~16) and predicting \textsf{lrPredict} (line:~22). 
The \textsf{lr} object wraps all the logical rules and required relations (e.g. parameters with default value 0) of the \datalog implementation for Logistic Regression.
When initializing \textsf{lr}, users can exploit the built-in functions to set the hyper-parameters that control the maximum number of iterations, the method used for  parameter initialization, and many others.  
After fitting the model to \textsf{df}, the \textsf{lrPredict} object could make predictions on the testing instances with the pre-trained model, \textsf{lrModel}. 
In both the fitting and predicting processes, the information of \datalog execution runtime can be obtained by using \textsf{session} as an input argument, which is same as the practice of Apache Spark.

\begin{figure}[h!t]
	\centering
	\includegraphics[width=0.4\textwidth]{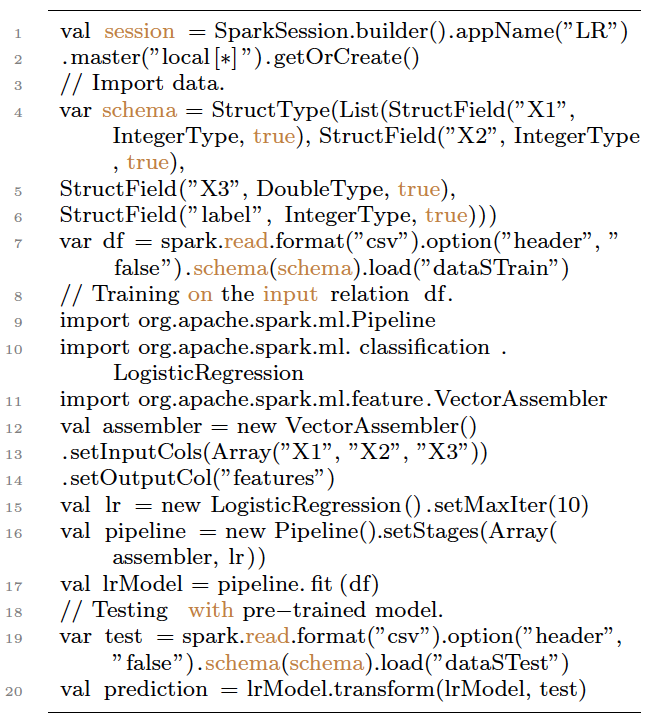}
	\vspace{-.5em}
	\caption{Snippet Code: Implementation with MLlib}\label{fig:mllib}\vspace{-1em}
\end{figure}

For the sake of comparison, we also show how Apache Spark \mlib will be used to implement the above example. 
The snippet code is shown in Figure~\ref{fig:mllib}. 
The pipeline of functionalities is very similar to that of our APIs; this will make it much easier using  the DataFrame APIs in our library for those who are already familiar with \mlib. 
Although there are minor differences in the aspects of data formatting and usage of some public functions, e.g. transform and assembler, the expression of \mlib and our \llib are very similar and both user-friendly.

\section{Demonstration}\label{sec-demo}

\begin{figure}[ht]
	\centering
	\includegraphics[width=0.4\textwidth]{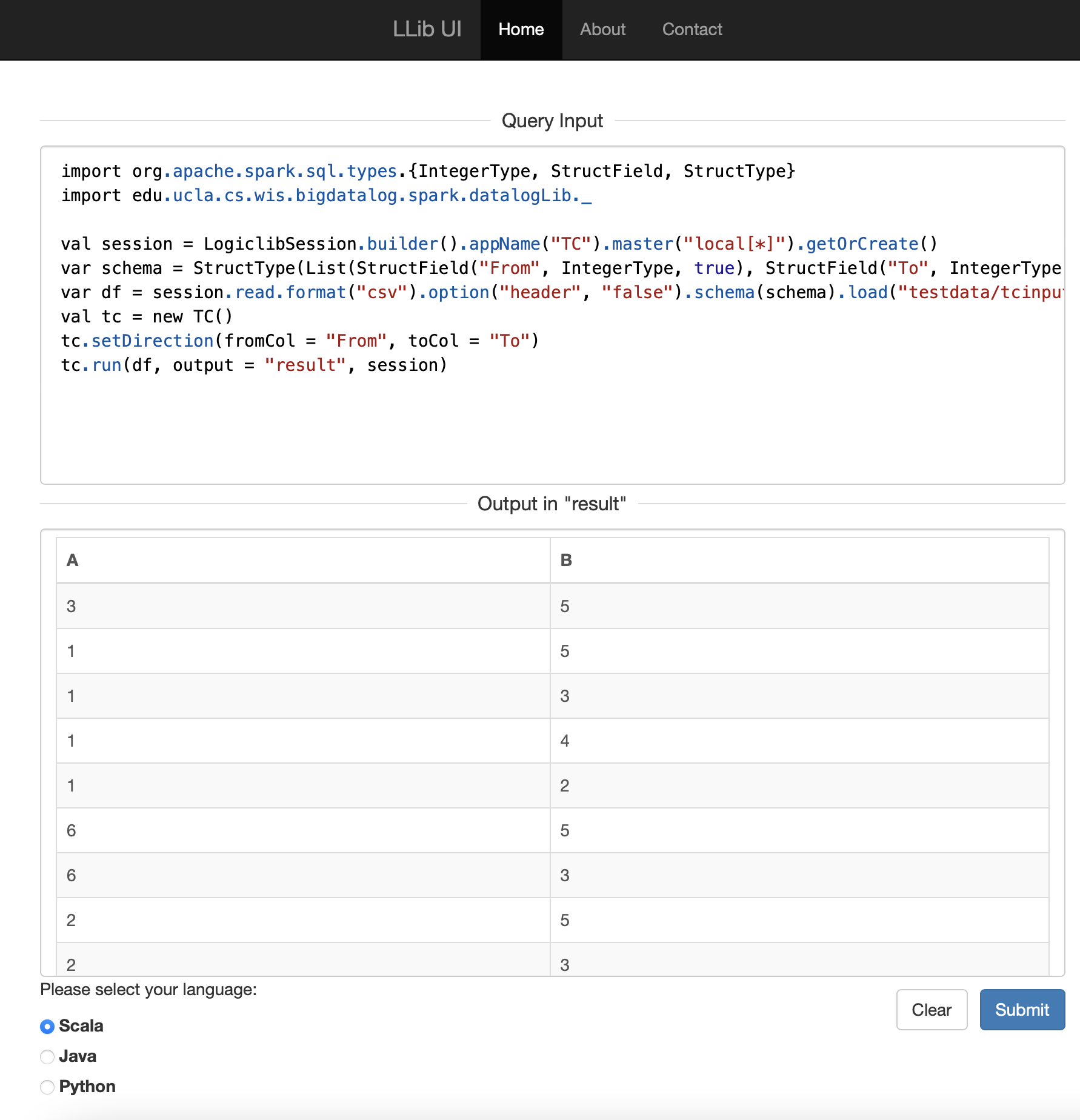}
	\vspace{-.5em}
	\caption{The User Interface of \llib}\label{fig:demo}
\end{figure}\vspace{-.5em}

Figure~\ref{fig:demo} illustrates the Web-based user interface of the \llib framework.
The upper part is a text box for users to type in the programs to be executed, while the lower part shows the results of execution.
For novice users who are not familiar with Apache Spark and Data Frame, we provide some pre-defined examples which can be reused by simply replacing the name of functions and the initialization of relations in the program. 
Advanced users are welcome to come up with the program by encapsulating new Datalog programs into user defined functions. 
We have imported all supported functions in the second line of the upper text box shown in Figure~\ref{fig:demo}.
Users can feel free to use other functionalities of DataFrame provided by Apache Spark.
We also support three kinds of programming languages in this demonstration: Python, Scala and Java. 
Users can switch the language by clicking the options in the bottom left.

Our plan of demonstration consists of two steps:

Firstly, we will give a brief introduction to the functionalities of \llib, in which we will exhibit the key components and features of our framework.
To this end, we will explain the detailed steps of initializing the session and schema for some simple graph queries, such as transitive closure and connected components.

Secondly, we will offer a hands-on step in which the public is invited to directly interact with the system and test its capabilities.
Users can write the programs with algorithms provided by \llib freely by themselves on different kinds of applications, including graph algorithm, machine learning  and data mining algorithms.
We also encourage audience to try different programming languages and implement the same program with \mlib so as to gain similar user experience with the official library in Apache Spark.

	\bibliographystyle{abbrv}
	\bibliography{llib}
	
\end{document}